\documentclass[aps,prb,twocolumn,amsmath,amssymb,superscriptaddress,10pt]{revtex4-2}
\usepackage[english]{babel}
\usepackage{graphicx}
\usepackage{bm}
\usepackage{amsmath,amssymb}
\makeatletter
\newsavebox\myboxA
\newsavebox\myboxB
\newlength\mylenA

\newcommand*\xoverline[2][0.75]{%
    \sbox{\myboxA}{$\m@th#2$}%
    \setbox\myboxB\null
    \ht\myboxB=\ht\myboxA%
    \dp\myboxB=\dp\myboxA%
    \wd\myboxB=#1\wd\myboxA
    \sbox\myboxB{$\m@th\overline{\copy\myboxB}$}
    \setlength\mylenA{\the\wd\myboxA}
    \addtolength\mylenA{-\the\wd\myboxB}%
    \ifdim\wd\myboxB<\wd\myboxA%
       \rlap{\hskip 0.5\mylenA\usebox\myboxB}{\usebox\myboxA}%
    \else
        \hskip -0.5\mylenA\rlap{\usebox\myboxA}{\hskip 0.5\mylenA\usebox\myboxB}%
    \fi}
\makeatother
\usepackage[export]{adjustbox}
\usepackage{float}
\usepackage{siunitx}
\usepackage{booktabs}
\usepackage{color}
\usepackage{xcolor}
\usepackage [latin1]{inputenc}
\usepackage{dcolumn}
\usepackage{etoolbox}
\usepackage{url}
\usepackage[colorlinks]{hyperref}
\usepackage[normalem]{ulem}
\usepackage{array}
\newcolumntype{x}[1]{>{\centering\arraybackslash\hspace{0pt}}p{#1}}
\usepackage{multirow}
\usepackage{accents}
\usepackage{braket}

\begin{document}

\title{Transverse and longitudinal magnons in strongly anisotropic antiferromagnet FePSe$_3$}

\author{F.~Le Mardel\'e}
\email[]{florian.le-mardele@lncmi.cnrs.fr} 
\affiliation{LNCMI-EMFL, CNRS UPR3228, Univ.\ Grenoble Alpes, Univ.\ Toulouse, Univ.\ Toulouse 3, INSA-T, Grenoble and Toulouse, France}

\author{A.~El Mendili}
\affiliation{Universit\'e Grenoble Alpes, CEA, IRIG, PHELIQS, 38000 Grenoble, France}

\author{M.~E.~Zhitomirsky}
\affiliation{Universit\'e Grenoble Alpes, CEA, IRIG, PHELIQS, 38000 Grenoble, France}
\affiliation{Institut Laue-Langevin,  71 avenue des Martyrs, 38042 Grenoble,  France}

\author{I.~Mohelsky}
\affiliation{LNCMI-EMFL, CNRS UPR3228, Univ.\ Grenoble Alpes, Univ.\ Toulouse, Univ.\ Toulouse 3, INSA-T, Grenoble and Toulouse, France}

\author{D.~Jana}
\affiliation{LNCMI-EMFL, CNRS UPR3228, Univ.\ Grenoble Alpes, Univ.\ Toulouse, Univ.\ Toulouse 3, INSA-T, Grenoble and Toulouse, France}

\author{I.~Plutnarova}
\affiliation{Department of Inorganic Chemistry, University of Chemistry and Technology Prague, 16628 Prague, Czech Republic}

\author{Z.~Sofer}
\affiliation{Department of Inorganic Chemistry, University of Chemistry and Technology Prague, 16628 Prague, Czech Republic}

\author{C.~Faugeras}
\affiliation{LNCMI-EMFL, CNRS UPR3228, Univ.\ Grenoble Alpes, Univ.\ Toulouse, Univ.\ Toulouse 3, INSA-T, Grenoble and Toulouse, France}

\author{M.~Potemski}
\affiliation{LNCMI-EMFL, CNRS UPR3228, Univ.\ Grenoble Alpes, Univ.\ Toulouse, Univ.\ Toulouse 3, INSA-T, Grenoble and Toulouse, France}
\affiliation{CENTERA Labs, Institute of High Pressure Physics, PAS, 01-142 Warsaw, Poland}

\author{M.~Orlita}
\email[]{milan.orlita@lncmi.cnrs.fr} 
\affiliation{LNCMI-EMFL, CNRS UPR3228, Univ.\ Grenoble Alpes, Univ.\ Toulouse, Univ.\ Toulouse 3, INSA-T, Grenoble and Toulouse, France}
\affiliation{Institute of Physics, Charles University, Ke Karlovu 5, Prague, 121 16 Czech Republic}
\date{\today}

\begin{abstract}
FePSe$_3$ is a collinear honeycomb antiferromagnet with an easy-axis anisotropy and large spins $S=2$. It belongs to a family of magnetic van der Waals materials, which recently attracted considerable attention. In this work we present an experimental magneto-optical study of the low-energy excitation spectrum in FePSe$_3$, together with its theoretical description. The observed response contains several types of magnon excitations. Two of them are conventional transverse magnons described by a classical theory of antiferromagnetic resonance. Two other modes are identified as multimagnon hexadecapole excitations with an anomalous $g$ factor approximately equal to four times the $g$ factor of a single Fe$^{2+}$ ion. These quasiparticles correspond to full reversals of iron spins that coherently propagate in the up-down antiferromagnetic structure.  They constitute  a novel type of collective excitations in anisotropic magnetic solids, called longitudinal magnons. Comparison between theory and experiment 
allows us to estimate the microscopic parameters of FePSe$_3$ including exchange coupling constants and the single-ion anisotropy. 
\end{abstract}
\maketitle

\section{Introduction}

\begin{figure}[t]
\includegraphics[width= .8\columnwidth,valign=t]{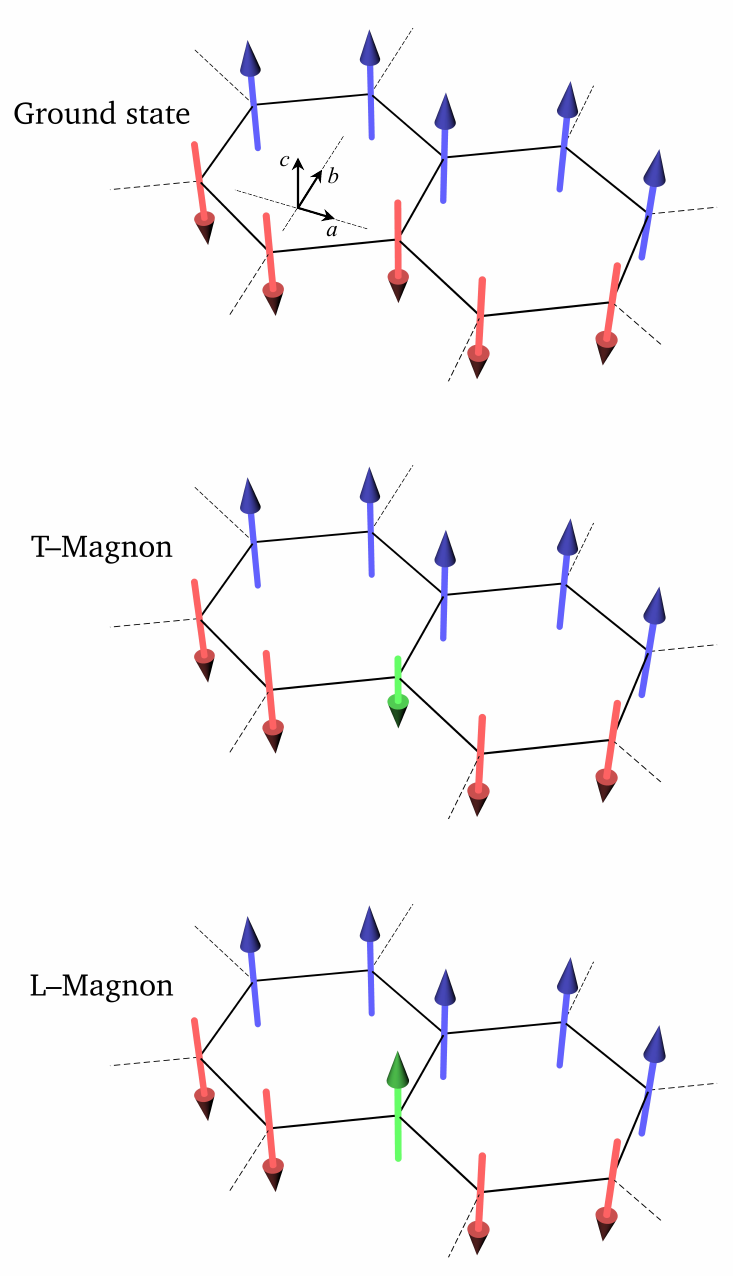} 
\caption{Cartoon representation of T- and L-magnons localized on a single lattice site (green arrow). Top panel: a collinear zigzag antiferromagnetic structure in FePSe$_3$ and FePS$_3$ materials. Middle panel: an ordinary transverse or T-magnon with  $|\Delta S^z|=1$. 
Bottom panel: a longitudinal or L-magnon with $|\Delta S^z|=2S$ ($=4$ for $S=2$ of Fe$^{2+}$ ions). 
}
\label{fig11}
\end{figure}

FePSe$_3$ belongs to a family of transition-metal (M) chalcogen-phosphates  with a generic chemical formula MPX$_3$ (X= S, Se). 
These materials offer a unique testbed for studying quasi two-dimensional  ferromagnetic and antiferromagnetic structures, in insulating or metallic 
phase, with bulk or few-layer samples down to the monolayer limit~\cite{Park16,SamalJMCA21,ChittariPRB16}. The electronic and magnetic 
properties of FePSe$_3$ are similar to those of the more studied sister compound FePS$_3$. Both materials are semiconductors with energy 
band gaps in the near-infrared range. Magnetic Fe$^{2+}$ ions have a large spin $S=2$ and form a honeycomb lattice in the $ab$ plane \cite{Taylor:1974aa,Flem:1982,Li:2019aa}. Stacking of honeycomb layers follows the rhombohedral symmetry in FePSe$_3$, whereas in FePS$_3$ adjacent layers are slightly shifted in accordance with the monoclinic crystal structure~\cite{Flem:1982}. Both materials order antiferromagnetically below comparable temperatures: $T_N = 118$~K  (FePS$_3$) and $T_N = 106$~K  (FePSe$_3$) \cite{Ferloni89}. In the ordered phases, their iron spins are oriented orthogonally to the $ab$ layers and form 
ferromagnetic zigzag chains in the $a$ direction that alternate antiferromagnetically along the $b$ axis \cite{WiedenmannSSC81,RulePRB07}, 
see Fig.~\ref{fig11}.

 The spin dynamics have been investigated in a greater detail for FePS$_3$. Specifically,  inelastic neutron-scattering measurements mapped the magnon dispersion in the entire Brillouin zone and provided us with values for the microscopic exchange interactions up to the third neighbors in the $ab$ plane~\cite{WildesJPCM12,LanconPRB16,WildesJAP20}. The magneto-optical response of FePS$_3$ in the ordered phase is dominated by an ``acoustic'' magnon with the gap $\Delta_1^s=15.1$~meV visible in Raman and infrared experiments~\cite{SekinePRB90,Wang2D16,McCrearyPRB20,LiuPRL21,VaclavkovaPRB21,Wyzula:2022}. In a magnetic field, this mode splits into two branches linear in $B$, with $S^z = \pm 1$ and the gyromagnetic factor of  $g\approx 2.15$. Thus, FePS$_3$ provides a textbook example of antiferromagnetic resonance (AFMR) in an easy-axis antiferromagnet~\cite{KittelPR51,KefferPR52,Kittel_book}. 
 
 The recent magneto-optical experiments on FePSe$_3$ found a new appealing effect in this material in comparison to FePS$_3$. The lowest antiferromagnetic (``acoustic'') magnon located at $\Delta_1^s \simeq 14.4$~meV interacts with a pair of quasi-degenerate chiral phonons~\cite{LuoNL23,Cui:2023}. The chirality selective hybridization couples the phonon angular momentum to the magnon dipole moment. Interestingly, this coupling was observed on FePSe$_3$ samples, using polarization-resolved magneto-Raman scattering technique,  
 down to a monolayer limit~\cite{LuoNL23}. 

The magnetic excitation spectrum of FePS$_3$ exhibits an unusual richness at high energies~\cite{Wyzula:2022}. Apart from the acoustic magnon, there is an ``optical'' magnon mode with $\Delta_1^a=39.6$~meV and same $g$ factor. The presence of two pairs of $k=0$ magnons is a direct consequence of the 4-sublattice antiferromagnetic structure. The corresponding gaps are in good agreement with the neutron-scattering experiments  \cite{LanconPRB16,WildesJAP20}. Still, at higher energies,
there is an extra line at $\Delta_4=57.5$~meV, which splits in magnetic field with a large effective $g$ factor: $g_4\approx 4g$. This specific magnetic mode was interpreted~\cite{Wyzula:2022} in terms of a full reversal of an iron spin. Such an excitation carries a large angular momentum, $|S^z|= 2S= 4$, and can be labelled as a longitudinal (L) magnon in order to distinguish it from conventional transverse (T) magnons with $|S^z|= 1$, corresponding to transverse precession of coupled magnetic dipoles.

Similar excitations were theoretically discussed for spin-1 easy-axis ferro- and antiferromagnets \cite{SilberglittPRB70,WierschemPRB12} and experimentally observed in FeI$_2$ with effective $S=1$ \cite{Petit80,KatsumataPRB00,BaiNP21}. For $S=1$, the angular momentum of an excitation with a reversed spin is  
$|S^z|=2$, hence, it was called a single-ion 2-magnon bound state \cite{SilberglittPRB70}. For FePS$_3$ with $S=2$, the L-magnon can be optionally called a single-ion 4-magnon bound state. In order to distinguish this type of excitations from the exchange-driven multi-magnon states observed in recent experiments \cite{DallyPRL20,LegrosPRL22}, we keep the term L-magnon, see also Sec.~\ref{Theory}. 
 
In this article, we report on the infrared magneto-spectroscopy study of FePSe$_3$ in a broader frequency range compared to the previous works on this material \cite{LuoNL23,Cui:2023,JanaPRB23}. Similarly to FePS$_3$, we find an ``optical'' magnon at $\Delta_1^a\approx 35$~meV. In addition, we observe {\em two pairs} of L-magnons with energies $\Delta_4^s\approx 53$~meV and $\Delta_4^a\approx 55$~meV. The presence of two pairs of L-magnons implies that these excitations have a sizeable dispersion in the Brillouin zone rather than being completely localized on the same site as the classical excitations in Ising models do. Thus, we extend and strengthen the previous observation of L-magnons with $S^z=4$ in FePS$_3$~ \cite{Wyzula:2022} by demonstrating their collective nature and firmly establishing them as a novel type of quantum quasiparticles. 

The paper is organized as follows. Section~\ref{Theory} provides a general theoretical description of L-magnons. The obtained experimental results are described in Sec.~\ref{Experiments}. By comparing theory and experiment for two pairs of T- and L-magnons,  we are able to extract the microscopic parameters of FePSe$_3$, which may stimulate future neutron-scattering studies of this material.

\section{Theory}
\label{Theory}

\subsection{Transverse magnons}

The two-dimensional spin Hamiltonian for layered FePX$_3$ (X = S, Se) materials with the easy-axis anisotropy 
can be written as \cite{LanconPRB16,Wyzula:2022}
\begin{equation}
\hat{\cal H }=\sum_{\langle i j\rangle } J_{ij} \mathbf{S}_i\cdot\mathbf{S}_j
  - D\sum_i{S^z_i}^2 - g\mu_BB\sum_iS^z_i \,.
\label{H0}
\end{equation}
Here,  $\mathbf{S}_i$ are $S=2$ operators residing on a honeycomb lattice. Exchange interactions $J_{ij}$ up to the third neighbors are necessary to include in order to describe the zigzag antiferromagnetic structure~\cite{Rastelli79,WildesPRB20}. We neglect magnetic coupling between planes, which is two orders of magnitude weaker than the largest in-plane exchange coupling~\cite{LanconPRB16,PawbakeACSNano22}. On the other hand, the single-ion constant $D$ is comparable to the in-plane exchanges producing a pronounced magnetic anisotropy in the magnetization of FePS$_3$ \cite{WildesPRB20} and large excitation gaps.

In collinear antiferromagnets, magnons are coherently propagating spin flips with $S^z = \pm 1$, see Fig.~\ref{fig11}. At the semiclassical level, such excitations can be viewed as transverse precession of coupled magnetic dipoles.
The standard spin-wave theory based on the Holstein-Primakoff representation of spin operators is commonly used to compute the magnon dispersion. The full expressions for magnon bands in FePX$_3$ are given, for example, by Wildes {\it et al.}~\cite{WildesJPCM12}. 
Below we provide only energies of the $k=0$ magnons that are directly measured in the (magneto-)optical experiments. For the up-up-down-down magnetic structure of FePX$_3$, the spectrum consists of two pairs of T-magnons corresponding to symmetric ($s$) and antisymmetric ($a$) oscillations of parallel spins in ferromagnetic zigzag chains:
\begin{equation}
\Delta_{1}^s=2S\sqrt{\xoverline[0.7]{D}(\xoverline[0.7]{D}+J_1+4J_2+3J_3)}\pm g\mu_BB
\label{transverse_magnon_sym}
\end{equation}
and 
\begin{equation}
\Delta_{1}^a=2S\sqrt{(\xoverline[0.7]{D}-2J_1+4J_2)(\xoverline[0.7]{D}-J_1+3J_3)}\pm g\mu_BB\,.
\label{transverse_magnon_antisym}
\end{equation}
In these expressions we have included a renormalized  constant $\xoverline[0.7]{D} = D[1-1/(2S)]$  \cite{Rastelli75,Balucani79,Oitmaa08}, which merely reflects the fact that a single-ion anisotropy is mute for $S=1/2$. Formally,  this quantum correction corresponds to the next order in the $1/S$ expansion and is often neglected in the linear spin-wave fits of the experimental data, but see \cite{Wheeler09} for an exception. As a result, the bare 
single-ion anisotropy constant deduced, {\it e.g.}, from the AFMR measurements is underestimated: for $S=2$ by $\sim 30$\%. In the following theoretical description of L-magnons, one has to employ a bare parameter $D$. For that we straightforwardly modify the previously reported anisotropy values to 
include this correction. 

\subsection{Longitudinal magnons}

Two-magnon bound states appear due to the attraction between spin flips on adjacent sites in ferromagnets \cite{Wortis63,Hanus63} as well as in materials with mixed ferro- and antiferromagnetic bonds \cite{Chubukov91}. 
The	$n$-magnon bound states with $n>2$ can be also created by the same mechanism, but dimensionality and frustration are crucial for their 
presence \cite{Kecke07}. Alternatively, for $S > 1/2$, a single-ion anisotropy of the easy-axis sign induces attraction for spin flips on the same site \cite{SilberglittPRB70}. It is energetically favorable to bring together as many spin flips as allowed by the spin quantum number $S$. The corresponding $2S$-magnon bound states can emerge without presence of bound states in the lower-magnon subsectors. Because of this and in order to avoid an unnecessary dependence of terminology on the quantum spin value, we call them longitudinal or L-magnons as opposed to ordinary transverse or T-magnons.
Longitudinal magnons may appear as a distinct type of magnetic excitations in materials with a strong easy-axis anisotropy. 
As anisotropy weakens, the L-magnon band starts to overlap with the $2S$-magnon continuum, and acquires a finite lifetime due to magnon-magnon interaction.

A convenient starting point for a theoretical description of L-magnons is the large-$D$ limit, $D\gg J_{ij}$. 
A local excitation $\ket{\pm S}\to\ket{\mp S}$ acquires energy from broken exchange bonds with adjacent spins but leaves
unchanged the single-ion term. Keeping a full spin flip completely localized, one can easily compute its energy \cite{KatsumataPRB00,Wyzula:2022}. For the zigzag antiferromagnetic structure of FePX$_3$ materials this yields: 
\begin{equation}
\Delta_4 = 2S^2(-J_1+2J_2+3J_3)\pm 4g\mu_BB \,.
\label{D4}
\end{equation}
In this `Ising' approximation, the energy of L-magnons has no $k$-dependence and the excitations remain completely localized. The dispersion of L-magnons and, hence, their 
collective nature is recovered by performing an expansion in powers of $J_{ij}/D$. Specifically, we associate two degenerate states $(\ket{+S},\ket{-S})_i$
on every site with a pseudo-spin-1/2 doublet $(\ket{\uparrow},\ket{\downarrow})_i$ and consider separately expansion for each exchange bond.
For general spins $S>1/2$, one has to resort to at least the $2S$ order to realize a flip $\ket{\pm S}\to\ket{\mp S}$. 
This results in the effective bond Hamiltonian:
\begin{equation}
\hat{\mathcal{H}}^{\rm eff}_n = \Tilde{J}^\perp_n\left( s_i^x s_j^x + s_i^y s_j^y\right)+\Tilde{J}^z_n s_i^z s_j^z
+ {\rm const} \ ,
\label{Heff}
\end{equation}
where $s^\alpha_i$ are spin-1/2 operators.

For FePX$_3$ materials with $S=2$, the transverse and the Ising constants in Eq.~(\ref{Heff}) are expressed as
\begin{equation}
\label{Jeff}
\Tilde{J}^\perp_n= - \frac{J_n^4}{4D^3}+\frac{3J_n^5}{8D^4}\,,\ \  \Tilde{J}^z_n=16J_n+\frac{4J_n^2}{3D}. 
\end{equation}
Here, we include two leading contributions for each of $\Tilde{J}_n$ from an infinite series in powers of $J_n/D$. The higher orders additionally generate
multisite terms in the effective Hamiltonian. 

An L-magnon corresponds to a flip of a pseudo spin  $\ket{\uparrow}\to\ket{\downarrow}$.
The Ising interactions between pseudo spins reproduce the expression (\ref{D4}) in the leading order,
whereas transverse terms are responsible for the L-magnon dispersion. The L-magnon band width is obviously small in comparison to that of the ordinary magnons, 
but still non-negligible. The full dispersion of L-magnons can be obtained by performing the harmonic spin-wave calculation for the effective spin-1/2 Hamiltonian 
combining all bond contributions (\ref{Heff}).
As a result, we obtain  two doubly-degenerate branches:
\begin{eqnarray}
{E_{\mathbf{k}}^\pm}^2 & = & \frac{1}{4} \bigl( A^2+|B_0|^2-B_1^2-|B_2|^2 \bigr)
\label{lm-dispersion} \\
&  \pm  & \frac{1}{2} \sqrt{\left|AB_0-B_1B_2\right|^2- [\operatorname{Im}(B_0B_2^*)]^2} 
\nonumber
\end{eqnarray}
with 
\begin{eqnarray}
&& A  =  3\Tilde{J}^z_3-\Tilde{J}^z_1+2[\Tilde{J}^z_2+\Tilde{J}^\perp_2\operatorname{Re}{(\gamma_1\gamma_3^*)}], 
\nonumber \\
&& B_0  =  (\gamma_1+\gamma_3)\Tilde{J}_1^\perp\,,\ \ B_1  =  \operatorname{Re}[\gamma_2^*(\gamma_1+\gamma_3)]{\Tilde{J}^\perp_2},
  \nonumber \\
&& B_2 =  \gamma_2\Tilde{J}^\perp_1+ \bigl({\gamma_1^*}^2+ {\gamma_2^*}^2 + {\gamma_3^*}^2\bigr)\Tilde{J}^\perp_3,
 \\
&& \gamma_1= e^{ik_x},\ \ \gamma_2= e^{i(-k_x+\sqrt{3}k_y)/2},\ \ \gamma_3= e^{-i(k_x+\sqrt{3}k_y)/2}.
\nonumber 
\end{eqnarray}
An illustration of this dispersion is presented below in Fig.~\ref{fig-lm} using microscopic parameters extracted later on from the 
experimental data for FePSe$_3$. Interestingly, the L-magnons have almost quasi-one-dimensional bands with a strongly
dispersive direction parallel to the ferromagnetic chains of the zigzag structure. This property follows naturally from small  
$\Tilde{J}_n^\perp \ll \Tilde{J}_n^z$. 

The energy gaps for the symmetric and antisymmetric L-magnons with $k=0$ measured in the optical experiments are 
expressed as
\begin{equation}
\label{lm-gaps}
\Delta_{4}^{s,a}=\Delta_4 -\frac{2J_1^2-4J_2^2-6J_3^2}{3D} + \Tilde{J_2}^\perp \mp |\Tilde{J}^\perp_1| \,,
\end{equation}
where $\Delta_4$ is given by Eq.~(\ref{D4}) with $S=2$.  The above expression is based on  the strong-coupling expansion, which
generally loses its accuracy for $J_n\sim D$. Still, from comparison of the first two contributions into the effective interaction 
parameters (\ref{Jeff}) we can conclude that 
the ``center-of-mass''  position of two modes $(\Delta_4^s + \Delta_4^a)/2$ is given by Eq.~(\ref{lm-gaps}) with a higher accuracy  
compared to the gap splitting $(\Delta_4^a -\Delta_4^s)$.

In  FePS$_3$, the lower L-magnon mode is found at $\Delta_4^s \approx 57.5$~meV and, thus, lies outside the 4-magnon continuum, which starts at $4\Delta_1^s = 60.4$~meV \cite{Wyzula:2022}, see Table~\ref{Tab}. The upper L-magnon 
mode $\Delta_4^a$ is nearly vanishing. Most likely, this mode merges with the onset of 4-magnon continuum, and therefore, decays quickly due to magnon-magnon interaction. We have used the set of microscopic parameters
provided for FePS$_3$  in \cite{WildesJAP20} (Table~\ref{Tab2}) to estimate from Eq.~(\ref{lm-gaps}) the gap splitting as
$\Delta_{4}^a-\Delta_{4}^s\approx 2.4$~meV. This value is sufficient to make $\Delta_4^a$ approach the 4-magnon continuum boundary and, thus, may explain its weak intensity. 

 \begin{figure*}[t]
	\includegraphics[width=1\linewidth]{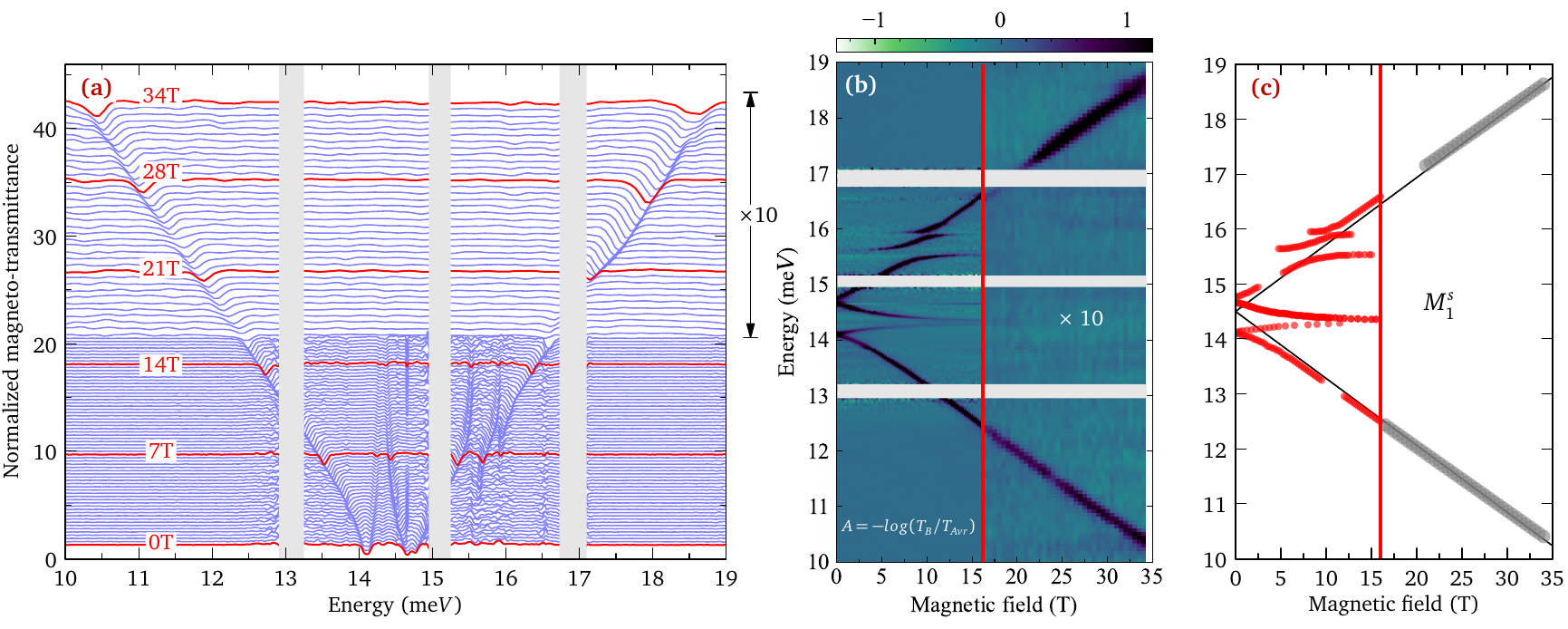}
\caption{Magneto-transmission data collected on FePSe$_3$ in the THz range at $T=4.2$~K plotted in the form of a stack-plot (a) and a false-color map (b). Two distinct sets of data are plotted, collected in the superconducting and resistive coils, below and above 16~T, respectively. Each magneto-transmission spectrum is normalized by the transmission averaged over the explored range (0-16T and 0-34T). The grey areas correspond to the region of full opacity due to optical phonon modes.
The panel (c) comprises the extracted magneto-transmission minima. Only the high-field part of the data (gray full circles) was used to fit (solid lines) the $\Delta_1^s$ energy and the corresponding $g$ factor (Tab.~\ref{Tab}).}
	\label{fig1}
\end{figure*}

\section{Experimental results}
\label{Experiments}

FePSe$_3$ was prepared by chemical vapor transport from elements in quartz glass ampoule. Iron (3N, 5-9~micron, Strem, Germany) phosphorus (6N, 2-6~mm granules, Wuhan Xinrong New Material Co., China) and selenium (6N, 2-6~mm granules, Wuhan Xinrong New Material Co., China) were placed in ampoule corresponding to 20~g of FePSe$_3$. Iodine was used as a transport medium (0.4~g, 3N, 1-3~mm, Fisher Scientific, USA) together with 1 at\% excess of selenium and phosphorus. The elements were sealed in a quartz ampoule under high vacuum ($1\times10^{-3}$~Pa, diffusion pump with a liquid nitrogen trap) using oxygen-hydrogen torch. The ampoule was first placed in a muffle furnace to react elements and form bulk FePSe$_3$. The ampoule was heated first at 450$^{\circ}$C for 25~h, subsequently to 500$^{\circ}$C for 50~h and on 600$^{\circ}$C for 50~h. The heating rate was 0.2$^{\circ}$C/min and cooling rate 1$^{\circ}$C/min. The ampoule with bulk FePSe$_3$ was placed in two zone furnace and first the growth zone was heated on 700$^{\circ}$C, while the source zone was kept at 500$^{\circ}$C. After 50 hours the thermal gradient was reversed and source zone was heated to 700$^{\circ}$C, while the growth zone temperature was decreased from 650$^{\circ}$C to 600$^{\circ}$C over a period of 5 days and following 5 days was kept at 600$^{\circ}$C. Finally during cooling procedure the growth zone was kept at 300$^{\circ}$C for 2 hours to remove all transport media from the growth zone. The ampoule was open in an argon-filled glovebox. 

The magneto-optical response of FePSe$_3$ was probed in the transmission mode and in the Faraday configuration, with the magnetic field perpendicular to the sample surface (along the $c$ axis). The radiation from a mercury lamp, or alternatively, a globar was analyzed by a Bruker Vertex 80v Fourier-transform spectrometer, and delivered to the sample via light-pipe optics. The FePSe$_3$ sample -- with an effective irradiated area of several mm$^2$ and thickness of $\approx$100~$\mu$m -- was kept at $T=4.2$~K in the helium exchange gas. In experiments using a superconducting coil (below 16~T), the radiation was detected by a composite bolometer placed right below the sample. When working with the resistive coil (above 16~T), to avoid excessive noise generated by the cooling water, the sample was placed on a mirror and probed in so-called double-pass transmission mode. The radiation was then detected by an external bolometer. The latter configuration is possible thanks to the insulating nature of the sample, thus staying highly transparent at low photon energies (apart from several phonon modes). 

 \begin{figure*}[t]
\includegraphics[width=1\linewidth]{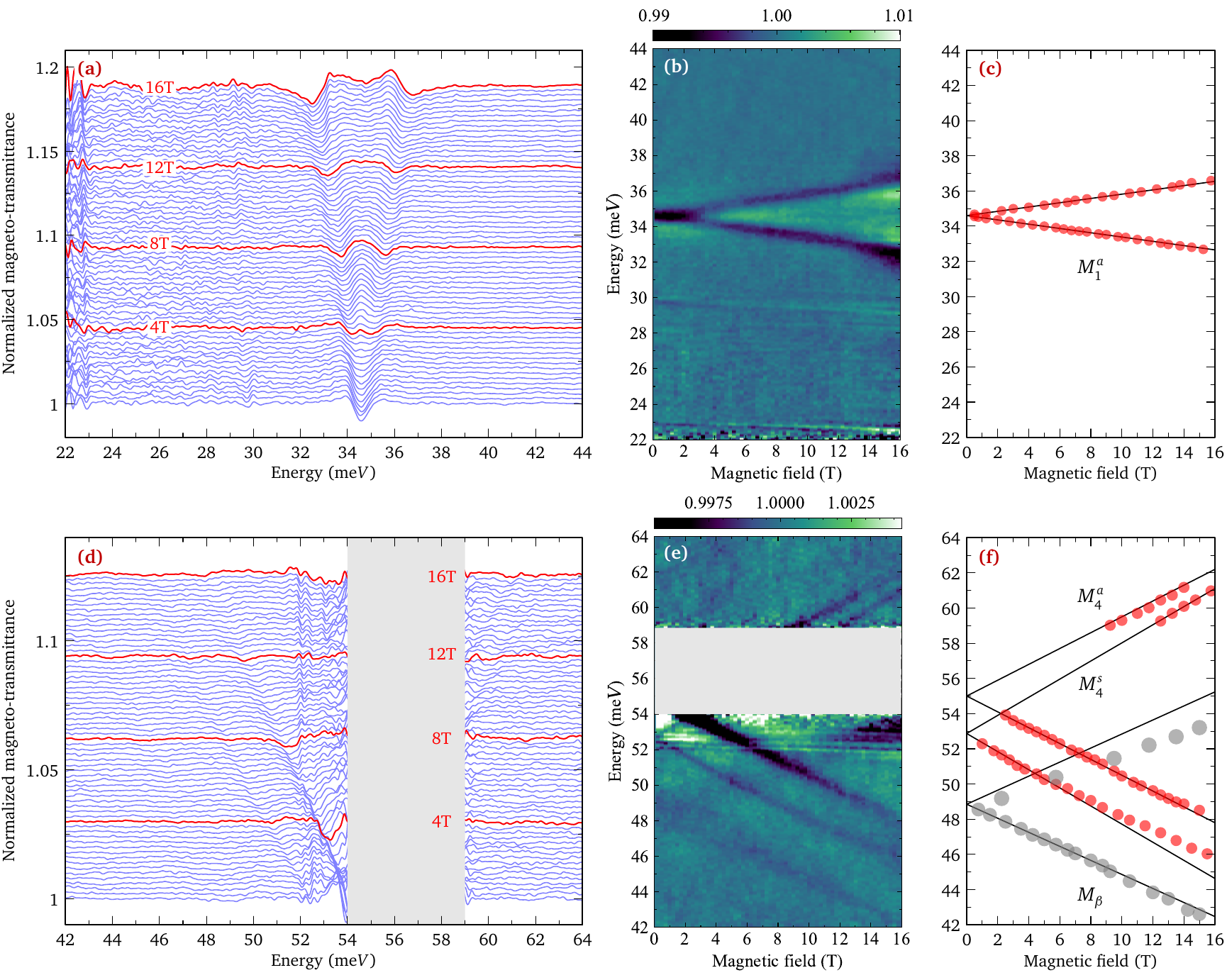}
\caption{Magneto-transmission data collected on FePSe$_3$ at $T=4.2$~K in two mid-infrared spectral ranges plotted in the form of a stack-plot (a,d) and a false-color map (b,e) in the magnetic field up to 16~T. In both cases, the magneto-transmission spectra are normalized by the transmission spectrum averaged over an interval of 10T centered at $B$. The grey areas correspond to the region of full opacity due to optical phonon modes. The panel (c) comprises the extracted magneto-transmission minima. The solid lines are fitted using the classical Kittel formula, yielding the zero-field energies of $\Delta_\beta$, $\Delta_4^s$ and $\Delta_4^a$ and the corresponding effective $g$ factors (Tab.~\ref{Tab}).}
\label{fig2}
\end{figure*}

The collected magneto-transmission data are plotted in Figs.~\ref{fig1} and \ref{fig2}. In the three selected spectral ranges, five distinct magnon-type excitations have been identified. A remarkable similarity has been found between the response of FePSe$_3$ and FePS$_3$ which greatly facilitates the interpretation of individual modes. To be consistent with Ref.~\cite{Wyzula:2022}, we have labelled these modes as M$^s_1$, M$^a_1$, M$_\beta$, M$^s_4$ and M$^a_4$ in order respecting the energy of transitions. The corresponding zero-field energies are $\Delta_1^s$, $\Delta_1^a$, $\Delta_\beta$, $\Delta_4^s$ and $\Delta_4^a$, respectively. The deduced energies are summarized in Tab.~\ref{Tab} and compared to those found in FePS$_3$. Let us now proceed with a more in-depth analysis. 

We start with the two most pronounced modes M$_1^s$ and M$^a_1$ having energies $\Delta_1^s=(14.5\pm0.5)$~meV and $\Delta_1^a=(34.6\pm0.2)$~meV, respectively. These two modes are conventional one-magnon gaps or -- using the above introduced terminology -- T-magnon modes. The upper ``optical'' mode M$^a_1$ follows the classical Kittel's formula for the AFMR~\cite{KittelPR51}: $\Delta_1^a(B) = \Delta_1^a\pm g_1 \mu_B B$ where the effective $g$ factor reaches 2.1. The same characteristic behavior is observed for the ``acoustic'' mode M$^s_1$ at higher magnetic fields (above 10~T). 

\begin{figure*}
\includegraphics[width=0.95\linewidth]{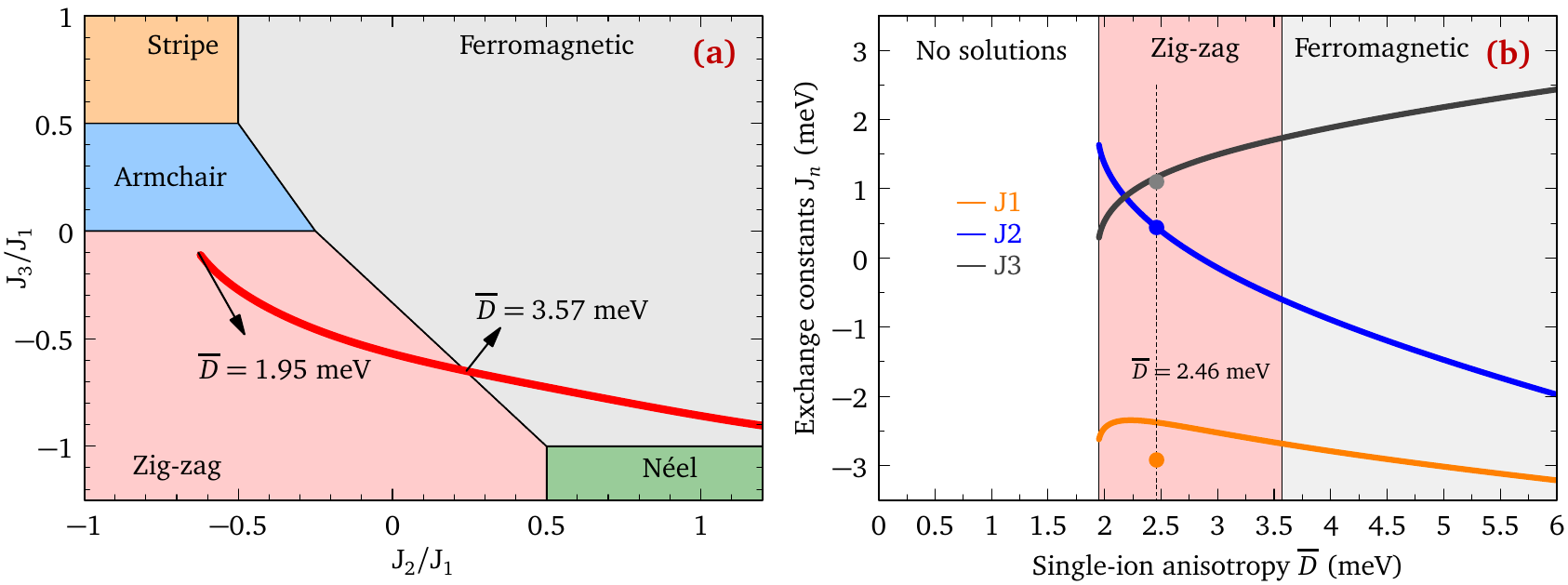}
\caption{(a) Phase diagram of the $J_1$--$J_2$--$J_3$ honeycomb antiferromagnet with strong easy-axis anisotropy, see the main text and Ref.~\cite{WildesPRB20}. The red curve represents solutions of Eqs.~(\ref{transverse_magnon_sym})--(\ref{D4}) parametrized using the single-ion constant 
$\xoverline[0.7]{D}$. No solution exists for $\xoverline[0.7]{D}<1.9$~meV. The zigzag antiferromagnetic structure appears only for $1.9<\xoverline[0.7]{D}(\mathrm{meV})<3.6$.   (b) Exchange coupling constants values $J_n$ for $n=1$--3 corresponding to the experimental gaps  as functions
of the anisotropy parameter $\xoverline[0.7]{D}$. Full circles in (b) show values of exchange constants in FePS$_3$ consistent with the L-magnon gap  in this material \cite{Wyzula:2022}.}
\label{fig3}
\end{figure*}

The M$^s_1$ mode exhibits, at low magnetic fields, a distinctively different behavior   from the classical AFMR. This is due to a pronounced magnon-phonon (magnon-polaron) coupling, already observed in the preceding infrared and Raman magneto-spectroscopy studies of FePSe$_3$~\cite{LuoNL23,Cui:2023,JanaPRB23}. This coupling was also found in the sister compound FePS$_3$~~\cite{VaclavkovaPRB21,LiuPRL21,ZhangCM21,PawbakeACSNano22} where the lower one-magnon gap has a nearly identical energy. To estimate the bare energy of $\Delta_1^s$, free of magnon-phonon coupling, we extrapolated the high-field part of the data using Kittel's formula~\cite{KittelPR51}. 

Other two well visible modes, M$^s_4$ and M$^a_4$, appear at higher photon energies, in the mid-infrared spectral range: $\Delta_4^s=(52.5\pm0.5)$~meV and $\Delta_4^a=(55.0\pm0.5)$~meV. At first glance, they also resemble the classical AFMR response -- both modes split symmetrically into two branches that are linear in $B$. However, a closer inspection reveals that the effective $g_4$ factor is roughly 4$\times$ larger than $g_1$. These transitions may be interpreted in terms of a single-ion 4-magnon bound state, \emph{i.e.}, L-magnon modes in an $S=2$ system. Such an excitation, at almost the same photon energy, was recently identified also in FePS$_3$~\cite{Wyzula:2022}.  
Similar to FePS$_3$, the L-magnon in FePSe$_3$ is a weak excitation,
with the integral strength by a factor of $10^3$ smaller as compared to the M$^s_1$ mode. 

The remaining magnon-type mode M$_\beta$ is a very weak excitation. This seems to be another multi-magnon mode. In this particular case, however, we may only speculate about its origin. Again, a similar line has also been revealed in the response of FePS$_3$ and interpreted as a 3-magnon bound state. The extracted effective $g$ factor $g_\beta=6.9$ supports such a view. 

The similarity between the magneto-optical responses of FePS$_3$ and of FePSe$_3$ greatly facilitated the interpretation of magnon-like excitations observed in the latter material (cf. Figs.~\ref{fig1} and \ref{fig2} with~\cite{Wyzula:2022}). This similarity concerns the energies, effective $g$ factors, as well as the relative strength of individual modes. 
All this suggests that the theoretical model for magnon excitations~in FePS$_3$ -- invoked in~Ref.~\cite{Wyzula:2022} and extended in Sec.~\ref{Theory} -- is also relevant to FePSe$_3$.  

Since the energy separation of the observed longitudinal modes is relatively small ($\Delta_4^a-\Delta_4^s\approx 2.4$~meV) and our theory (\ref{lm-gaps}) describes this splitting only perturbatively, let us put it aside for a while. 
Instead, we assume that $(\Delta_4^s+\Delta_4^a)/2\approx\Delta_4^a\approx\Delta_4^s\approx 54$~meV. This allows us to use Eqs.~(\ref{transverse_magnon_sym}) and (\ref{transverse_magnon_antisym}), together with the center-of-mass in (\ref{lm-gaps}), \emph{i.e.}, with $ \Tilde{J}_1^\perp$=0, to estimate the exchange coupling constants and the single-ion magnetic anisotropy in FePSe$_3$ based on the experimentally determined energies of magnon modes -- a piece of information still missing in the literature.

The zero-field energies of magnon excitations: (\ref{transverse_magnon_sym}), (\ref{transverse_magnon_antisym}) and (\ref{lm-gaps}), represent a set of three equations for 4 unknown parameters: $J_n$ for $n=1$--3 and $\xoverline[0.7]{D}$. To proceed with such an underdetermined problem, we choose to parametrize it in $\xoverline[0.7]{D}$, and search for additional constraints, such as the experimentally established magnetic structure of FePSe$_3$  in zero field \cite{WiedenmannSSC81}. Clearly, the characteristic zigzag chains do not 
emerge for arbitrary combinations of the parameters.   

\begin{table}[t]
    \centering
    \renewcommand{\arraystretch}{1.6}
    \begin{tabular}{|x{15mm}|x{15mm}x{15mm}|x{15mm}x{15mm}|}
 \hline
        & \multicolumn{2}{c|}{\textbf{FePSe$_3$}} & \multicolumn{2}{c|}{\textbf{FePS$_3$}}\\   
     Magnon & Energy&$g$ factor&Energy&$g$ factor \\
        mode & (meV)&&(meV)&\\
\hline
\hline
         M$_1^s$&  14.5 &   2.1&  15.1&   2.15\\
         M$_1^a$&34.6&2.1&39.6&2.15\\
         M$_{\beta}$&48.8&6.9&54.0&5.0\\
         M$_4^s$&52.6&8.9&57.5&9.2\\
         M$_4^a$&55.0&7.8&60.6&7.3\\
\hline   
        
    \end{tabular}
    \caption{Energies of selected magnon modes observed in FePSe$_3$ extracted from our magneto-transmission data (Fig.~\ref{fig1} and \ref{fig2}) compared to analogous modes active in the response of FePS$_3$. The latter parameters are taken from Ref.~\cite{WildesJAP20} and were used to reproduce the L-magnon mode in preceding study of FePS$_3$ by Wyzula~et~al.~\cite{Wyzula:2022}. All values in the table are in meV.}
    \label{Tab}
\end{table}

This is shown in Fig.~\ref{fig3}(a), where we plot the phase diagram established in~\cite{WildesPRB20}. This diagram was obtained considering a reduced Ising-type Hamiltonian, with out-of-plane $S=2$ spins organized in a honeycomb lattice. Such an approach is applicable in the case of a relatively strong easy-axis anisotropy, \emph{i.e.}, $D\gtrsim \mathrm{max}|J_n|$ for $n=1$--3. This assumption is justified \emph{a posteriori}. The red curve in Fig.~\ref{fig3}(a) is a result of our parametrization, showing the ratios of the exchange constants as a function of $\xoverline[0.7]{D}$. We infer that the antiferromagnetic zigzag arrangement of spins is only possible for certain values of the magnetic anisotropy: $1.9<\xoverline[0.7]{D}(\mathrm{meV})<3.6$. For $\xoverline[0.7]{D}$ too large, a ferromagnetic phase emerges. For $\xoverline[0.7]{D}$ too small, no real solution exists.   

Values of the exchange constants $J_1$, $J_2$, and $J_3$ evaluated as a function of the (renormalized) single-ion parameter $\xoverline[0.7]{D}$ are plotted in Fig.~\ref{fig3}(b) and listed in Tab.~\ref{Tab2}. In the relevant interval of $\xoverline[0.7]{D}=(2.8\pm0.9)$~meV, the nearest-neighbor coupling $J_1$ remains nearly constant, allowing us to set its value as $J_1 = -(2.5\pm 0.2)$~meV. The other two exchange constants evolve considerably with $\xoverline[0.7]{D}$. This leaves us with relatively large uncertainties of their values: $J_2 = (0.5\pm 1.0)$~meV and $J_3 = (1.0\pm 0.7)$~meV. Nevertheless, the simplified equation (\ref{D4}) for the L-magnon energy does not depend on $\xoverline[0.7]{D}$. This allows us, after establishing the value of  $J_1$,  to find a simple approximate relation in the leading order: $2J_2+3J_3\approx 4.2$~meV, which helps us to reduce the uncertainty in the $J_2$ and $J_3$ parameters. 

It is instructive to compare the values of the exchange constants estimated for FePSe$_3$ with those already known for the sibling compound FePS$_3$. At present, there are several sets of parameters proposed for the latter material \cite{LanconPRB16,Okuda83,KurosawaJPSJ83}. In Fig.~\ref{fig3}(b), the color-coded full circles show the values of $J_n$ at the corresponding single-ion magnetic anisotropy $\xoverline[0.7]{D}=2.46$~meV. These values were recently suggested in Ref.~\cite{WildesJAP20}, where the magnon dispersion obtained in the neutron-scattering experiments was fitted using the 
extended version of the spin Hamiltonian~(\ref{H0}), which includes a biquadratic exchange. Notably, these values lead to the correct estimate of the 4-magnon energy (\ref{D4}) in FePS$_3$ and are only slightly different from those extracted here for FePSe$_3$. Both sets are compared in Tab.~\ref{Tab2}.

\begin{figure}[t]
\includegraphics[width=0.95\columnwidth]{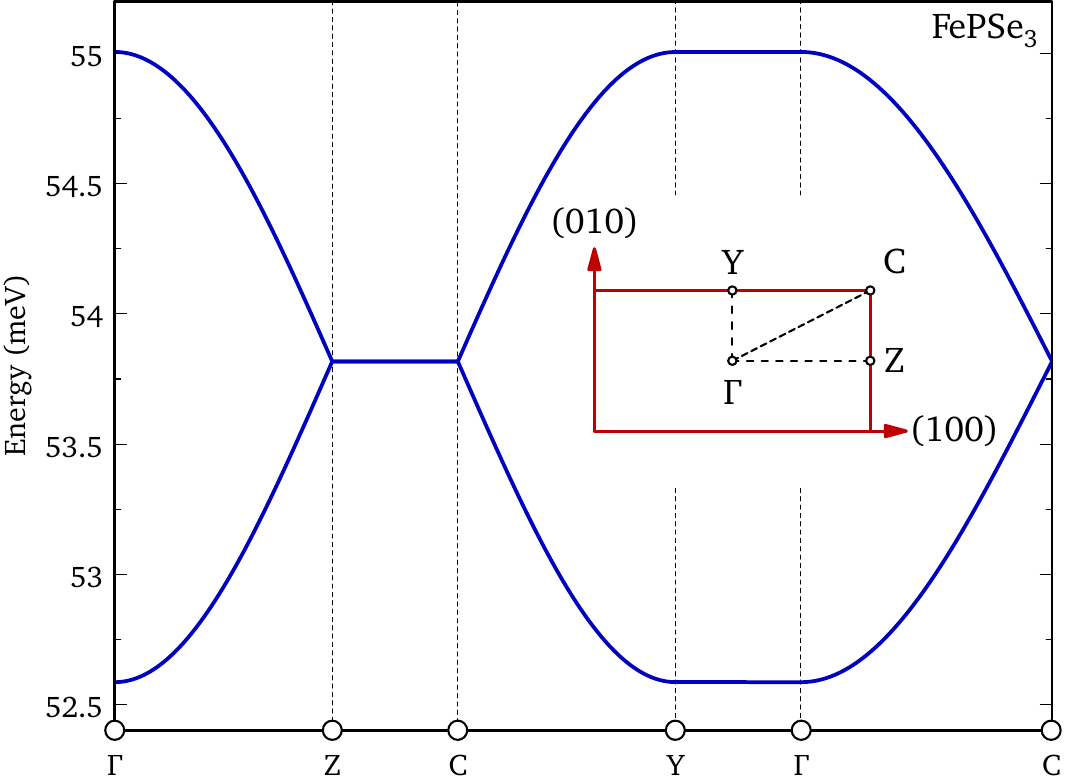}
\caption{Computed dispersion of longitudinal magnons using the microscopic parameters of FePSe$_3$ listed in Table~\ref{Tab2}.
The inset specifies the chosen path in the Brillouin zone. 
The most and least dispersive sections of the path correspond to parallel and orthogonal  directions with respect  to the ferromagnetic zigzag chains,
which are, in turn, parallel to the  crystallographic $a$  axis.
}
\label{fig-lm}
\end{figure}

Let us now return to the observed splitting of the longitudinal mode ($\Delta_4^s \neq \Delta_4^a$). According to Eq.~(\ref{lm-gaps}), two $k=0$ branches, the symmmetric and antisymmmetric one, are expected to appear for the L-magnon mode due to the 4-sublattice structure of FePSe$_3$, separated by an energy of $2| \Tilde{J}_1^\perp|$. Using Eq.~(\ref{Jeff}) and the above evaluated intervals for $J_1$ and $\xoverline[0.7]{D}$, we obtain for the splitting $\Delta_4^a-\Delta_4^s\in (0.2-7.7)$~meV. The experimentally observed value of 2.4~meV lies within this interval, thus corroborating our overall data interpretation. Let us also note that the two L-magnon modes identified in FePSe$_3$ display slightly different effective $g$ factors, see Tab.~\ref{Tab}. This difference suggests some additional spin-orbit effects, not accounted for in our simplified theoretical approach in Sec.~\ref{Theory}.

Figure~\ref{fig-lm} illustrates the in-plane dispersion of the L-magnon, with both the symmetric and antisymmetric branches that give rise to two optically active $k=0$ modes. As mentioned above, the plotted dispersion has almost a one-dimensional character, displaying a strongly dispersive direction parallel to the ferromagnetic chains of the zigzag structure. The dispersion was computed for the following set of parameters: $J_1=-2.47$~meV, $J_2=1.39$~meV, $J_3=0.5$~meV and $\xoverline[0.7]{D}=1.99$~meV. These values were obtained by using the full Eq.~(\ref{lm-gaps}) together with T-magnon energies from Eqs.~(\ref{transverse_magnon_sym}) and (\ref{transverse_magnon_antisym}). In such a case, we conveniently solve four independent equations and obtain four parameters as a result. Nevertheless, let us mention again that the splitting of the L-magnon modes in Eq.~(\ref{lm-gaps}) has a lower accuracy as compared to their mean (center-of-mass) energy and the exchange constants listed in Tab.~\ref{Tab2}. 

Finding this theoretically anticipated splitting in the experimental data serves as an indication that the observed 4-magnon line does not correspond to a localized state in the magnetic lattice, but instead, to a dispersing quantum quasi-particle. This finding also allows us to reinterpret the weak magnon-like mode $M_\alpha$ observed in our preceding study of FePS$_3$, see Supplementary materials of Ref.~\cite{Wyzula:2022}, which now appears to be the antisymmetric branch of the L-magnon $M_4^a$. This mode overlaps with the onset of the 4-magnon continuum, which explains its integral intensity, significantly weaker as compared to $M_4^s$ in FePS$_3$. 

\begin{table}[tb!]
    \centering
    \renewcommand{\arraystretch}{1.6}
    \begin{tabular}{|x{15mm}|x{15mm}|x{15mm}|}
 \hline
        & \textbf{FePSe$_3$} & {\textbf{FePS$_3$}}\\   
\hline
\hline
         $J_1$& -2.5 $\pm$ 0.2 &  -2.92 \\
         $J_2$&  0.5 $\pm$ 1.0  & 0.44\\
         $J_3$& 1.0 $\pm$ 0.7& 1.1\\
         $\xoverline[0.7]{D}$& 2.8 $\pm$ 0.9& 2.46\\
         $D$&3.7 $\pm$ 1.2& 3.28\\
\hline  
\end{tabular}
\caption{Microscopic parameters of FePSe$_3$ deduced from the magneto-optical measurements. Similar parameters obtained for FePS$_3$ 
\cite{WildesJAP20} are listed for comparison. All values are given in meV.}
    \label{Tab2}
\end{table}

\section*{Conclusions}

We have measured magneto-optically active  excitations in FePSe$_3$ in the infrared spectral range. The excitation spectrum is comprised of the conventional transverse magnons with $|S^z|=1$ and the longitudinal magnons with $|S^z|=4$. The latter can be viewed  as $2S$-magnon single-ion bound states brought up by a strong easy-axis anisotropy. The magneto-optical activity of these high-angular momentum excitations signify their mixed dipole-hexadecapole symmetry. Such a mixing may be induced by higher-order terms in the crystal-field Hamiltonian 
present due to the local $C_3$ rotation symmetry on magnetic sites of a honeycomb lattice or by other spin-orbital effects.
In contrast to the previously studied FePS$_3$ \cite{Wyzula:2022}, we have observed two split
pairs of L-magnons that correspond to symmetric and antisymmetric oscillations in the magnetic unit  cell, which contains four Fe$^{2+}$ ions. 
The strong-coupling  expansion ($D\gg J_{ij}$) was used to theoretically compute the observed splitting  and the overall dispersion 
of the L-magnon band. Investigation of the intermediate regime ($J_{ij}\sim D$), including L-magnon decay, is a subject of interest for our future work.

Comparison of the theoretical results with the experimental data 
allowed us to estimate the exchange parameters and the magnetic anisotropy constant for FePSe$_3$.
It will be  interesting to directly measure the L-magnon excitations together with their dispersion
 in the inelastic neutron-scattering experiments.  This would establish longitudinal magnons as a novel type of collective magnetic excitations 
 in strongly anisotropic magnetic solids. Finally, a finite band width  ($\sim 2.4~$meV) of the longitudinal magnons suggests that the condensation 
 of these $|S^z|=4$ quasiparticles at strong magnetic fields may result in a new quantum hexadecapole state stable in a few-Tesla range in the vicinity
of the condensation field. Such a possibility calls for high magnetic field measurements on FePSe$_3$ similar to the study performed on the sister material FePS$_3$ \cite{WildesPRB20}.

\begin{acknowledgments}
We thank A. R. Wildes for numerous discussions.
The work has been supported by the EU Graphene Flagship project. 
M.E.Z.\ acknowledges support by the ANR,  France,  Grant No.~ANR-19-CE30-0040.
The work was supported by the Czech Science Foundation, project No. 22-21974S.
Z.S. was supported by ERC-CZ program (project LL2101) from Ministry of Education Youth and Sports (MEYS) and used large infrastructure from project reg. No. CZ.02.1.01/0.0/0.0/15\texttt{\_}003/0000444 financed by the EFRR.
The authors also acknowledge the
support of the LNCMI-CNRS in Grenoble, a member of
the European Magnetic Field Laboratory (EMFL).
\end{acknowledgments}

\bibliography{FePSe3}

\end{document}